# Multielemental single-atom-thick *A* layers in nanolaminated V$_2$(Sn, *A*)C (*A*=Fe, Co, Ni, Mn) for tailoring magnetic properties

Youbing Li [1, 2, †], Jun Lu [3, †], Mian Li [1], Keke Chang [1], Xianhu Zha [1, 4], Yiming Zhang [1], Ke Chen [1], Per O. Å. Persson [3], Lars Hultman [3], Per Eklund [3], Shiyu Du [1], Zhifang Chai [1], Zhengren Huang [1], Qing Huang [1*]

[1] Engineering Laboratory of Advanced Energy Materials, Ningbo Institute of Materials Technology and Engineering, Chinese Academy of Sciences, Ningbo, Zhejiang 315201, China

[2] University of Chinese Academy of Sciences, 19 A Yuquan Rd, Shijingshan District, Beijing 100049, China

[3] Thin Film Physics Division, Department of Physics, Chemistry, and Biology (IFM), Linköping University, SE-581 83 Linköping, Sweden

[4] Center for Quantum Computing, Peng Cheng Laboratory, Shenzhen 518055, China

**Abstract**

Tailoring of individual single-atom-thick layers in nanolaminated materials offers atomic-level control over material properties. Nonetheless, multielement alloying in individual atomic layers in nanolaminates is largely unexplored. Here, we report a series of inherently nanolaminated $V_2(A'_xSn_{1-x})C$ (A'=Fe, Co, Ni and Mn, and combinations thereof, with x≈1/3) synthesized by an alloy-guided reaction. The simultaneous occupancy of the four magnetic elements and Sn, the individual single-atom-thick A layers in the compound constitute high-entropy-alloy analogues, two-dimensional in the sense that the alloying exclusively occurs in the A layers. $V_2(A'_xSn_{1-x})C$ exhibit distinct ferromagnetic behavior that can be compositionally tailored from the multielement A-layer alloying. This two-dimensional alloying provides a structural-design route with expanded chemical space for discovering materials and exploit properties.

Tailoring of individual single-atom-thick layers in nanolaminated materials offers atomic-level control over tailoring a desired property of a material. For example, artificially nanolaminated magnetic materials are widely applied in storage media and devices. Notably, the demonstration of the giant magnetoresistance (GMR) effect has enabled hard-drives and other storage media.[1] The GMR sensitivity is expected to be highest when only single-atom layers of ferromagnetic materials are sandwiched, since alignment of magnetization vectors of neighboring ferromagnetic atoms would then have the lowest energy cost.

Conceptually, this type of structure with single-atom-thick layers can be correlated to $M_{n+1}AX_n$ phases, which are a family of inherently nanolaminated ternary compounds with hexagonal crystal structure (space group *$P_{63}/mmc$*, *194*), where M is an early transition metal, A is mainly an element from groups 13–16, X is carbon and/or nitrogen, and n = 1-3.[2,3] Their crystal structure can be depicted by alternative stacking of $M_{n+1}X_n$ sublayer and a single-atomic layer of A, usually referred to as 211, 312 and 413 phases according to the value of n. If alloying or replacement of ferromagnetic elements at specific crystal sites in MAX phases could be realized, their magnetic properties may also be tailored. The key notion here is that the A layer is just one atomic layer thick. However, the few known magnetic MAX phases are based on alloying of Mn at the M site, such as $(Cr_{0.75}Mn_{0.25})_2GeC$,[4] $Mn_2GaC$,[5] $(V,Mn)_3GaC_2$,[6] $(Mo_{0.5}Mn_{0.5})_2GaC$,[7] $(Cr_{0.5}Mn_{0.5})_2GaC$,[8] and $(Cr_{0.5}Mn_{0.5})_2AuC$.[9] Fe-containing MAX phase materials were recently reported, specifically $Cr_{2-x}Fe_xGeC$,[10] $(Cr,Fe)_2AlC$,[11] and $Mo_2(GaAuFe)C$.[12]

Control of the occupancy of magnetic elements on the A rather than M sites would be crucial to tune magnetic properties. The finding of iron in the A site of $Mo_2(GaAuFe)C$ is encouraging, however, the presence of secondary iron-containing impurity phases in the $Mo_2(GaAuFe)C$ films impedes the determination of magnetic properties on the A plane of interest. Moreover, theoretical predictions suggest that Ni and Co should tend to occupy the M-sites of the MAX phases,[13] but this has not been demonstrated experimentally, likely due to the thermodynamically preferred formation of competing binary MA alloys or intermetallic phases. Generally, the late transition elements of Fe, Co, Ni, and Mn have not been considered as possible A-site elements in the definition of MAX phases. It would be of great interest to introduce these magnetic elements at A sites, since the overlapping between electron clouds of M and A atoms is limited compared to that between M and X atoms, which should aid in preserving the magnetic properties. A possible approach to realize this is by alloying with other main-group elements in such a way that the alloying neither transforms the atomic stacking in the crystal nor their bonding with the parent structure. Fe, Co, and Ni have been employed as additives for formation of $M'_3SnC_2$ (M′=Ti, Zr, Hf) phases via dissolution-diffusion-precipitation in intermediate $Z_xSn_y$ (Z= Fe, Co, Ni) alloys,[14,15] while none of Fe/Co/Ni were detected in the final MAX phases. The $A'_xSn_y$ (A′=Fe, Co, Ni, and Mn, or their binary mixtures) phases belong to the same space group ($P_{63}/mmc$ or $194$) as MAX phases. This similarity in crystal structures would facilitate the nucleation of MAX phases in saturated alloys by a reaction between a binary carbide and an intermediate state of $A'_xSn_y$ alloys. During such a reaction, these complex atoms in molten (or solid) alloys may thermodynamically and coherently arrange with [$M_6C$] octahedral building blocks to form ternary layer structure. Thus, it would be of great interest to obtain magnetic MAX phases by an alloy-guided reaction.

Here, we demonstrate this approach to A-site alloying of Sn with Fe/Co/Ni/Mn magnetic elements to synthesize a series of magnetic MAX phases of $V_2(A'_xSn_{1-x})C$ (A′=Fe, Co, Ni and Mn, or their binary/ternary/quaternary combination). The regular $M_2AX$ crystal structure of $V_2SnC$ is retained, with a mixture of two to five elements (Sn plus one or more of Fe, Co, Ni, and Mn) randomly occupying the A sites. The magnetic properties of $V_2(A'_xSn_{1-x})C$ MAX phases exhibit distinct ferromagnetic behavior that can be tailored by their constitutive elements. These results have wide-ranging implications since they not only can be used for tailoring magnetic properties, but most importantly demonstrate the formation of multielement A layers, as an analogy to high-entropy alloys[16-18], two-dimensional in the sense that alloying exclusively occurs in the single-atom-thick A layers, in layered transition metal carbides.

**MAX phases containing one magnetic element**

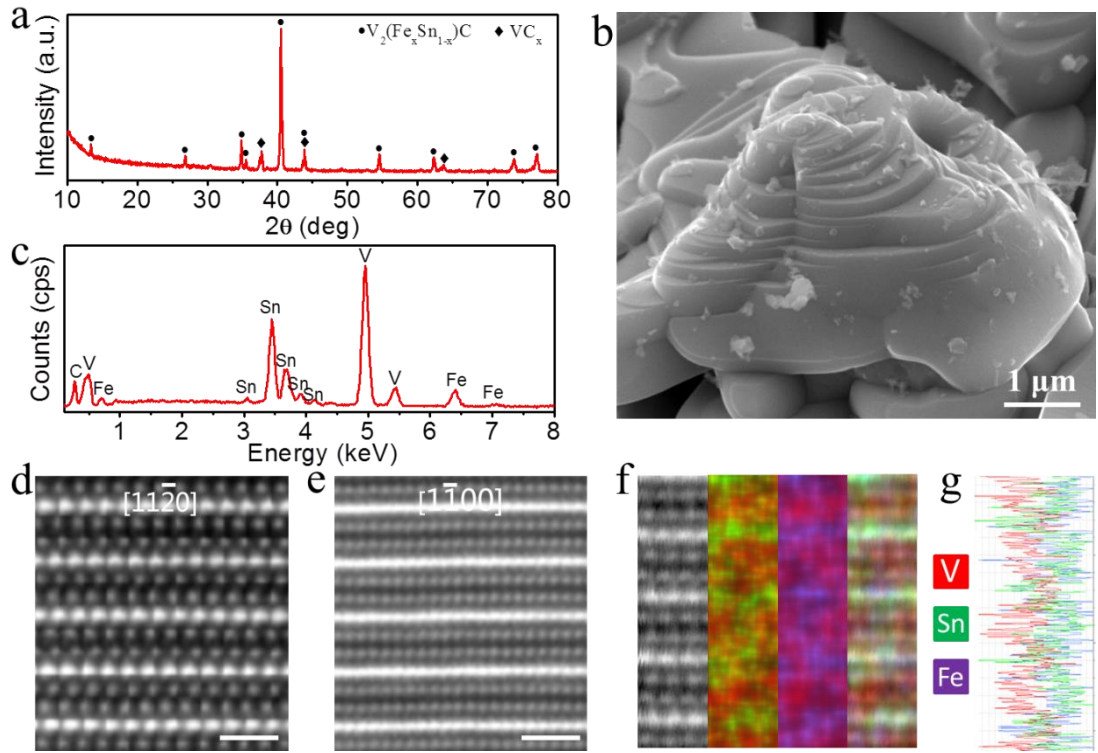

**Figure 1.** (a) XRD patterns of $V_2(Fe_xSn_{1-x})C$ after acid treatment. (b) SEM image of $V_2(Fe_xSn_{1-x})C$ and the corresponding energy-dispersive spectroscopy (EDS) analysis (c). High-resolution (HR)-STEM images of $V_2(Fe_xSn_{1-x})C$ showing atomic positions along [11$\bar{2}$0] (d) and [1$\bar{1}$00] (e) direction, respectively. (f) STEM-EDS mapping of V-Kα (red), Fe-Kα (purple) and Sn-Kα (green) signals, respectively. The STEM image scale bars are 1 nm.

Figure 1a shows an XRD pattern of the V-Fe-Sn-C system synthesized at 1100°C by molten salt method. The diffraction peaks are close to the characteristic crystal structure of $Cr_2AlC$ and $V_2AlC$ phases,[19,20] with peaks at 2θ≈13°, 2θ≈26° and 2θ≈40°, indicating the formation of a 211 MAX phase. Some byproducts, such as Sn metal, intermetallic compounds (FeSn$_2$) and non-stoichiometric vanadium carbide (VC$_x$), are present when the reaction temperature is 1000°C or 1100°C (Figure S1a). At temperatures of 1200°C or higher, the as-synthesized MAX phase tended to decompose due to out-diffusion of the A element (Figure S1a). This indicates the formation the previously not reported MAX phase V$_2$SnC. The microstructure of observed particles with terraces is typical of MAX phases (Figure 1b). In previously reported M′$_3$SnC$_2$ (M′=Ti, Zr, Hf) phases,[14] Fe was used as additive to aid the reaction, but absent in the final products. Here, however, the corresponding EDS spectra of these particles showed the presence of Fe, besides V, Sn and C (Figure 1c). The relative atomic ratio of V:(Fe+Sn) is approximately 2:1, consistent with the stoichiometry of 211 MAX phases (Table S1). The high content of Fe (9.4 at.%) and its uniform distribution in elemental mapping (Figure S2a) further indicate that Fe is incorporated in the as-synthesized MAX phase.

To determine the position of Fe, we performed high-resolution high-angle annular dark field scanning transmission electron microscopy (HAADF-STEM) and lattice-resolved energy dispersive X-ray (EDS) spectroscopy, as shown in Figure 1d-1g. STEM images acquired with the beam along the [11$\bar{2}$0] and [1$\bar{1}$00] zone axes are shown in Figure 1d and 1e, respectively. One layer of brighter spots (the A atomic layers) are interleaved by two adjacent layers of darker spots (the M atomic layers). Carbon is typically not visible because of its low contrast compared to the heavier M and A atoms, the characteristic zig-zag stacking of the M$_{n+1}$X$_n$ slabs viewed along the [11$\bar{2}$0] zone axes is seen.[21-23] STEM-EDS mapping and line-scan analysis are shown in Figure 1f and Figure 1g. V is in the M sites following the zig-zag stacking. Fe (purple) overlaps Sn (green), proving that both are in the A-site positions. The composition (EDS, Table S2), the composition of the elements V, Sn, and Fe is 67, 22, and 11 at.%, respectively. The molar ratio of V:(Sn+Fe) is about 2:1 and Sn:Fe is ~2:1, which is in good agreement with the above EDS result in SEM. Thus, the chemical formula of the resultant MAX phase is close to V$_2$(Sn$_{2/3}$Fe$_{1/3}$)C.

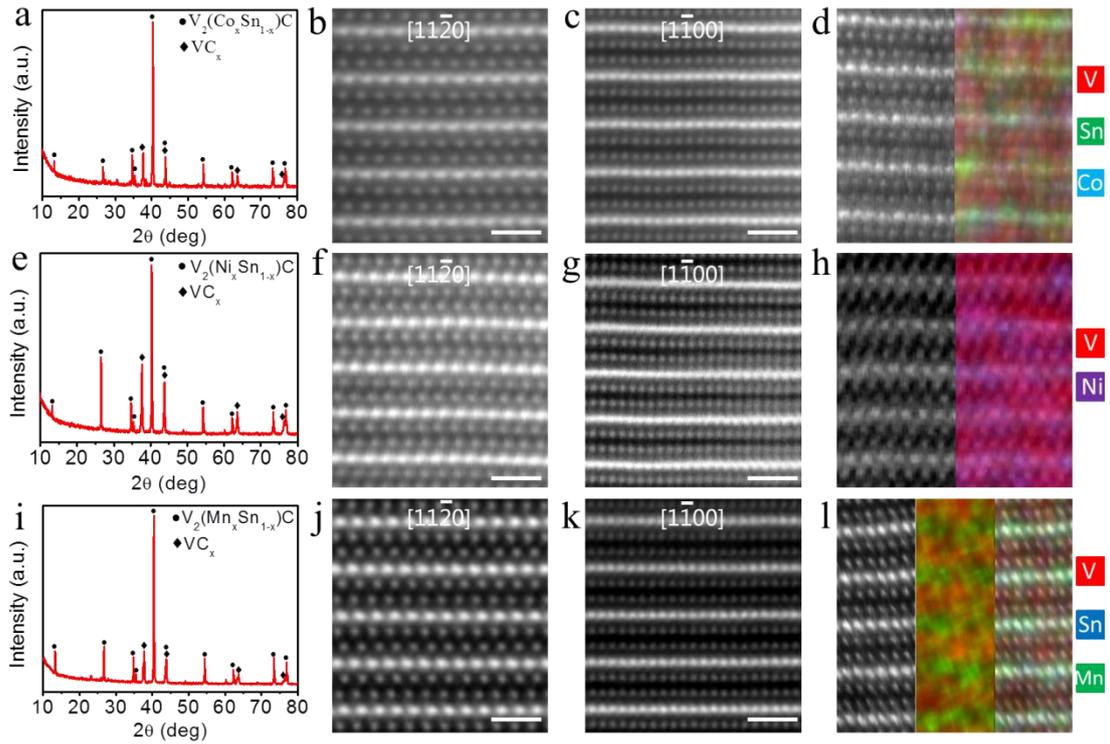

**Figure 2.** XRD patterns of $V_2(Co_xSn_{1-x})C$ (a), $V_2(Ni_{1/3}Sn_{2/3})C$ (e), and $V_2(Mn_{1/3}Sn_{2/3})C$ (i) after acid treatment. HR-STEM images of $V_2(Co_{1/3}Sn_{2/3})C$, $V_2(Ni_{1/3}Sn_{2/3})C$ and $V_2(Mn_{1/3}Sn_{2/3})C$ showing atomic positions along $[11\bar{2}0]$ (b, f, j) and $[1\bar{1}00]$ (c, g, k) direction, respectively. (d) STEM-EDS mapping of $V_2(Co_xSn_{1-x})C$ (d), $V_2(Ni_{1/3}Sn_{2/3})C$ (h), and $V_2(Mn_{1/3}Sn_{2/3})C$ (l) MAX phase. The STEM image scale bars are 1 nm.

To prove the generality of this methodology, we used Co, Ni, and Mn instead of Fe in the starting materials and followed the same chemical synthesis process. Figure 2 shows structural and chemical characterization (XRD, lattice-resolved STEM, and EDS in STEM) of the resulting products. Similar to $V_2(Fe_{1/3}Sn_{2/3})C$, the XRD patterns (Figure 2a, 2e, 2i, and Figure S1b-d) and SEM-EDS results (Figure S2b-d, Figure S3 and Table S1) of final products showed the formation of MAX phases and the incorporation of Co, Ni and Mn. STEM and atomically resolved EDS confirm the atomic position of these elements at the A site, mixed with Sn for $V_2(Co_{1/3}Sn_{2/3})C$ (Figure 2b-2d), $V_2(Ni_{1/3}Sn_{2/3})C$ (Figure 2f-2h), and $V_2(Mn_{1/3}Sn_{2/3})C$ (Figure 2j-2l). Atomically resolved elemental mapping (Figure 2d, 2h, and 2l) and line-scan analysis in STEM mode is also recorded (Figure S4). The molar ratio of Sn to the magnetic element in all these phases is close to 2:1, as in the Fe case.

## MAX phases containing two magnetic elements

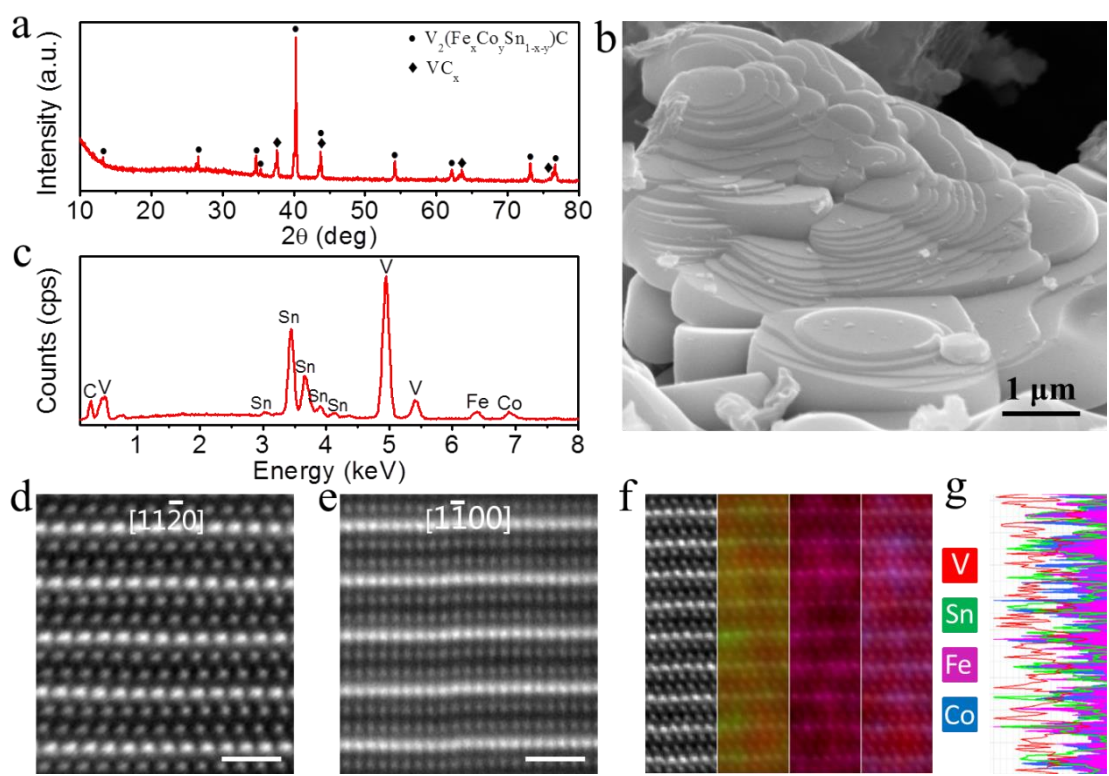

**Figure 3.** (a) XRD pattern of $V_2(Fe_{1/6}Co_{1/6}Sn_{2/3})C$ after acid treatment. (b) SEM image of $V_2(Fe_{1/6}Co_{1/6}Sn_{2/3})C$ and the corresponding energy-dispersive spectroscopy (EDS) analysis (c). HR-STEM images of $V_2(Fe_{1/6}Co_{1/6}Sn_{2/3})C$ showing atomic positions along $[11\bar{2}0]$ (d) and $[1\bar{1}00]$ (e) direction, respectively. (f) STEM-EDS mapping of V-K$\alpha$ (red), Fe-K$\alpha$ (pink), Co-K$\alpha$ (blue) and Sn-K$\alpha$ (green) signals, respectively. The STEM image scale bars are 1 nm.

If two magnetic elements can be simultaneously incorporated into the MAX structure, this would offer additional prospects for tuning magnetic properties, since strong spin-electron-coupling or interaction between different magnetic elements may enhance the magnetic properties, as known from, *e.g*, permalloy $Ni_{80}Fe_{20}$ with high magnetic permeability.[24] For co-incorporation of Fe and Co, the XRD pattern showed that the main product is a 211 MAX phase with small amounts of byproducts of Sn, intermetallic phases ($FeSn_2$, $CoSn_2$) and $VC_x$ (Figure 3a, Figure S5). SEM demonstrated the typical terraced laminate microstructure (Figure 3b). From the EDS results in SEM (Figure 3c), V, Fe, Co, Sn, and C five elements were detected in a layer-structure particle, and their relative molar proportions are 49.8, 4.2, 4.2, 16.7, and 26.1 at.%, respectively (Table S3). Elemental mapping of these five elements further provided evidence that Fe, Co and Sn have the same distribution (Figure S6a). The molar ratio of V:(Fe+Co+Sn) is very close to 2:1, and the Sn : (Co+Fe) ratio is 2:1. Therefore, the obtained MAX phase has the formula of $V_2(Fe_{1/6}Co_{1/6}Sn_{2/3})C$. Furthermore, STEM images of the $V_2(Fe_{1/6}Co_{1/6}Sn_{2/3})C$ phase with the beam along the [11$\bar{2}$0] and [1$\bar{1}$00] zone axes are shown in Figures 3d and 3e, respectively. The $V_2C$ sublayers exhibits the characteristic zig-zag pattern, like $V_2(A'_xSn_{1-x})C$ (A'=Fe, Co, Ni or Mn) MAX phases. No atomic disorder or phase separation was observed, despite the complex composition (three elements in the A layer). Further, atomically resolved EDS mapping analysis (Figure 3f) and line-scan analysis (Figure 3g) in STEM mode provided direct evidence that Fe, Co, and Sn elements all occupy A sites. None of these elements were detected at M site.

Following the same synthesis methodology, we also synthesized several other MAX phases with two magnetic elements on the A site, *i.e.*, $V_2(Fe_xNi_ySn_{1-x-y})C$, $V_2(Co_xNi_ySn_{1-x-y})C$, $V_2(Mn_xFe_ySn_{1-x-y})C$, $V_2(Mn_xCo_ySn_{1-x-y})C$, and $V_2(Mn_xNi_ySn_{1-x-y})C$. Phase identification, microstructures and elemental composition by XRD, SEM and EDS are presented in the Supplementary information (Figures S6, S7, S8, and S9 as well as Table S3).

**Multielement A-site phases**

A multielemental feature in chemical composition of a single-phase homogeneous material can generally have drastic implications for physical and chemical properties, as realized for so-called high-entropy alloys or multi-principal element alloys.[16-18] In analogy with this materials-design strategy, we posed the hypothesis that three or even four magnetic elements could be simultaneously incorporated since all these four (Fe, Co, Ni, and Mn) elements have similar electronegativity, atomic radii and itinerary electron number. We therefore performed the same synthesis for all combinations of three of these elements (Figure S10-S13). In all cases, a 211 MAX phase was the main phase, with small amounts of intermetallic phases and vanadium carbides, as shown by XRD (Figure S10a, S11a, S12a, and S13a). The corresponding EDS spectra and elemental mapping images in typical layer-structure particles identified multi-element composition of the final MAX phases (Figure S10b-S10d, S11b-S11d, S12b-S12d, and S13b-S13d). The compositions were also determined (Table S4), and resulting formula can be stated as $V_2(A'_xSn_{1-x})C$, where $A'$ is a combination of three elements from Fe, Co, Ni, and Mn, and $x \approx 1/3$.

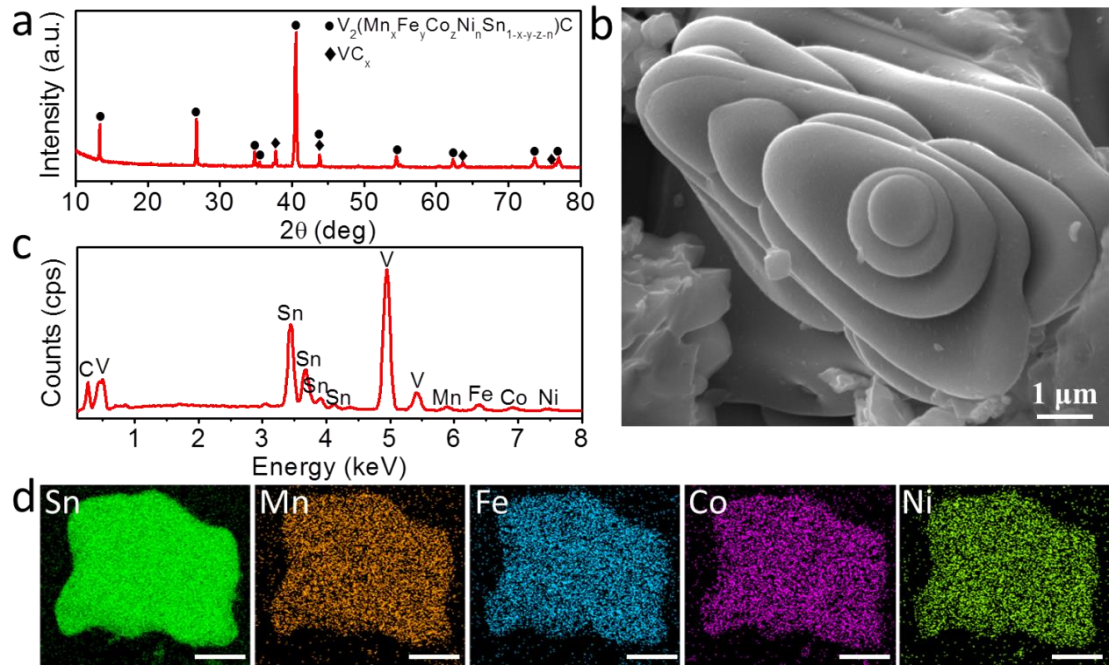

**Figure 4.** (a) XRD pattern of product showing character diffraction of $V_2(A'Sn)C$ ($A'$=Fe,Co,Ni,Mn). (b) The SEM image showing typical laminated structure of $V_2(A'Sn)C$ ($A'$= Fe,Co,Ni,Mn) and (c) corresponding energy-dispersive spectroscopy (EDS) analysis indicating the presence of Fe, Co, Ni, and Mn elements in products. (d) Elemental mapping on one particle clearing proving the uniform distribution of Fe, Co, Ni, Mn, and Sn. The map scale bars are 5 μm.

Furthermore, we synthesized a MAX phase with all four magnetic elements as well as Sn simultaneously. The XRD pattern shows that the final product is comprised of MAX phase and various tin alloys (Figure 4a, Figure S14). The laminated morphology of the particles is similar the above-mentioned MAX phases (such as in Figure 1b and 3b), but with more rounded edges (Figure 4b). EDS in SEM detected all constitutive elements (V, Sn, Fe, Co, Ni, Mn, and C) in these particles (Figure 4c), and the molar ratio of all four magnetic elements to Sn was close to 1:2 (Table S4). Elemental mapping of Sn, Fe, Co, Ni, and Mn corroborated that all these five elements have the same distribution (Figure 4d), and certainly occupy the A site. The obtained $V_2(A'_xSn_{1-x})C$ (where again A′ is a combination of Fe, Co, Ni, and Mn, and x is close to 1/3) thus constitutes a realization of a multielement (analogous to a high-entropy alloy) nanolaminated material, two-dimensional in the sense that the multielement alloying exclusively occurs on the A layers.

**Phase diagrams and stability**

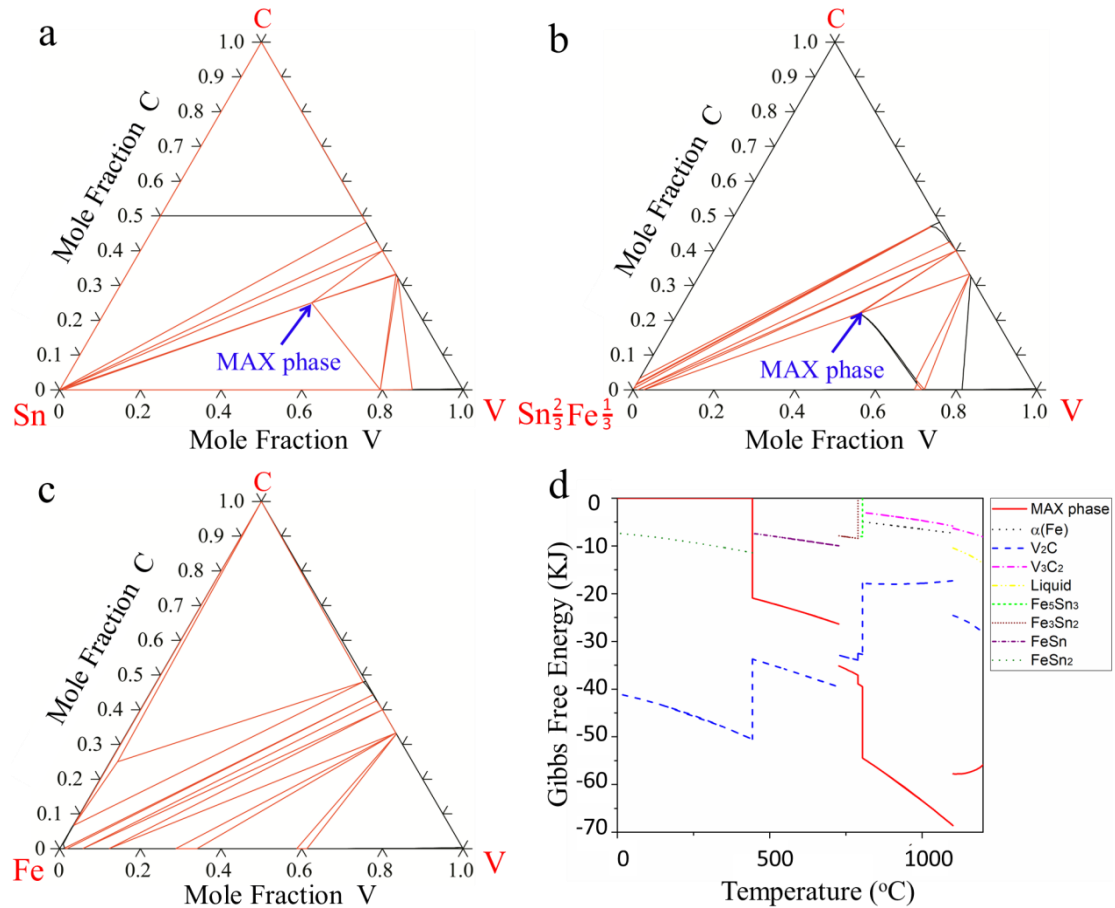

**Figure 5.** The isothermal sections at 1100 °C for phase diagrams of (a) V-Sn-C system, (b) V-(Sn,Fe)-C system and (c) V-Fe-C system. (d) Gibbs free energy of phases during experimental synthesis of $V_2$(Fe,Sn)C MAX phase.

Figure 5 shows the calculated isothermal sections at 1100 °C of the phase diagrams of V-Sn-C, V-($Sn_{2/3}Fe_{1/3}$)-C and V-Fe-C system. These results indicate that $V_2SnC$ is the only thermodynamically stable ternary phase in the V-Sn-C system (Figure 5a). In the case of Fe addition (Figure 5b), the $V_2(Fe_{1/3}Sn_{2/3})C$ phase can also be at equilibrium with $V_3C_2$ and $FeSn_2$, consistent with the experimental results (Figure 1a). However, in the V-Fe-C phase diagram, the hypothetical ternary MAX phase $V_2FeC$ is not stable (Figure 5c). Instead, vanadium carbides ($V_2C$ and $V_3C_2$) and a Fe-rich V-Fe intermetallic phase are the most competitive phases, corroborated by the experimental results. In fact, all our attempts to synthesize $V_2A'C$ (A'=Fe, Co, Ni, and Mn; *i.e.*, without Sn) phases failed (Figure S15).

The Gibbs free energy values of $V_2(Fe_{1/3}Sn_{2/3})C$ as well as intermediate phases during synthesis are shown in Figure 5d. According to this, $V_2(Fe_{1/3}Sn_{2/3})C$ can appear at temperature as low as 400°C. Below this temperature, the $V_2C$ and $Fe_5Sn_3$ phases are dominant. Above 750 °C, $V_2C$, and $Fe_5Sn_3$ gradually disappear and transform into $V_2(Fe_{1/3}Sn_{2/3})C$ phase by a peritectic reaction, *i.e.*, solid $V_2C$ and an intermediate liquid $Fe_5Sn_3$ transform into $V_2(Fe_{1/3}Sn_{2/3})C$. Without the presence of Fe (neither Co, Ni, and Mn), the formation of $V_5Sn_3$ is instead thermodynamically favored. Fe has higher affinity to Sn and thus a stronger tendency to form Fe-Sn alloys than V does. This should favor nucleation of $VC_{1-x}$ at low temperature and promote the peritectic reaction between $VC_{1-x}$ and liquid $Fe_5Sn_3$ alloy to form the final $V_2(Fe_{1/3}Sn_{2/3})C$ phase.

In general, the stable $V_2(A'_xSn_{1-x})C$ (A'= Co, Ni or Mn) MAX phases follow similar reaction paths as $V_2(Fe_{1/3}Sn_{2/3})C$, because of the reduced Gibbs free energy of the phase in the V-A'-Sn-C system through the addition of A' elements. The same is apparently true for multielement MAX phases $V_2(A_xSn_{1-x})C$, where A is two, three or four of Fe, Co, Ni, and Mn. The mixing entropy at the A site must therefore account for most of the decrease in Gibbs free energy and the corresponding thermodynamic stability. More exhaustive results and discussion of the stability of these phases, including density functional theory calculations, is provided in the Supplementary Information, Sections S4-S5.

**Magnetic Properties**

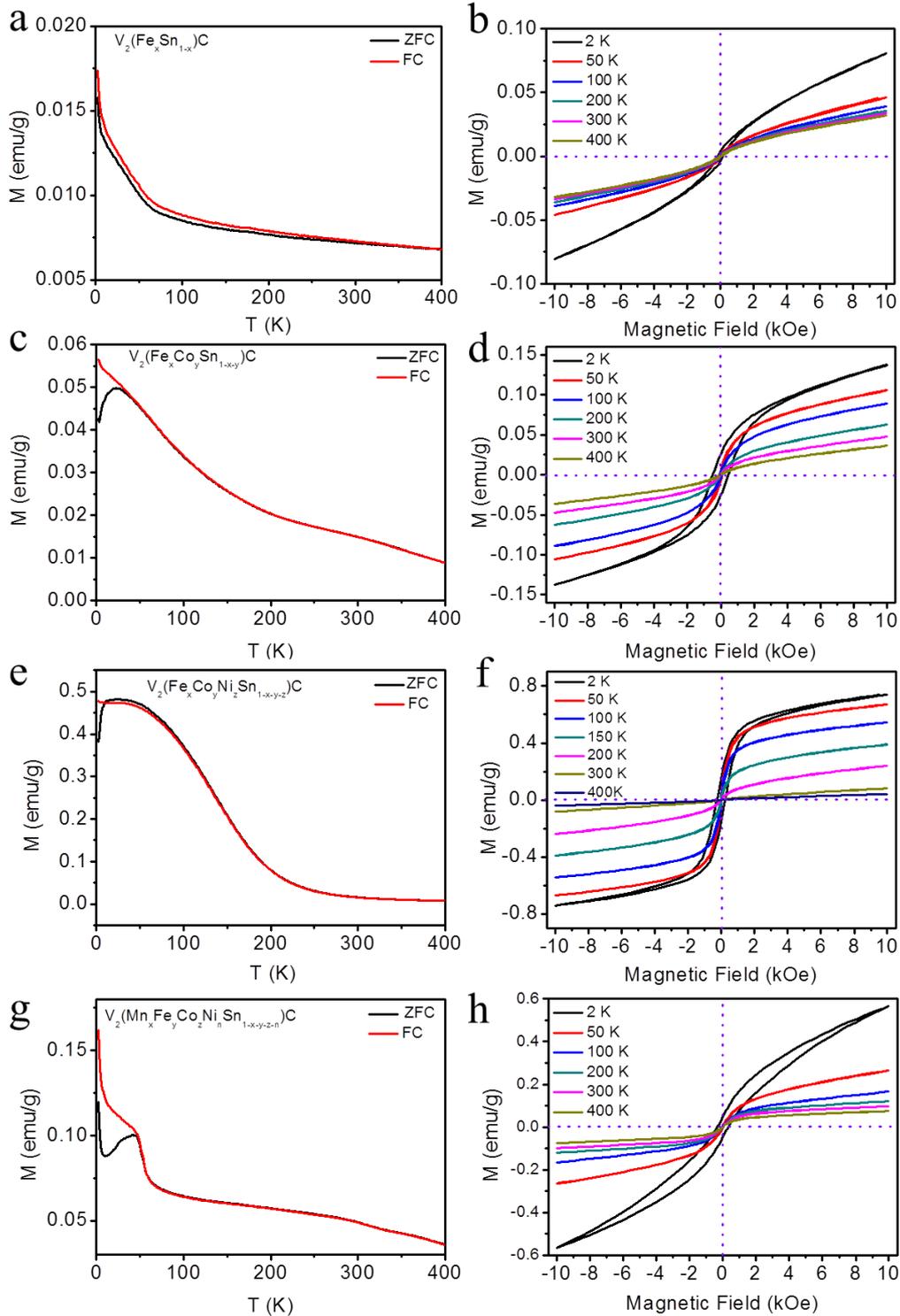

**Figure 6.** Temperature dependent magnetization M-T curves for the $V_2(Fe_xSn_{1-x})C$ (a)

$V_2(Fe_xCo_ySn_{1-x-y})C$ (c), $V_2(Fe_xCo_yNi_zSn_{1-x-y-z})C$ (e) and $V_2(Mn_xFe_yCo_zNi_nSn_{1-x-y-z-n})C$ (g) at 1000 Oe in the range of 2-400 K, respectively. Magnetic hysteresis loops of $V_2(Fe_xSn_{1-x})C$ (b), $V_2(Fe_xCo_ySn_{1-x-y})C$ (d), $V_2(Fe_xCo_yNi_zSn_{1-x-y-z})C$ (f), and $V_2(Mn_xFe_yCo_zNi_nSn_{1-x-y-z-n})C$ (h) at different temperature in the range of -10 kOe - 10 kOe, respectively.

The as-synthesized series of $V_2(A'_xSn_{1-x})C$ (A'=Fe, Co, Ni, and Mn, or their combination) was washed in acid to remove any Fe(Co,Mn,Ni)-containing secondary phases for subsequent magnetic measurements, to exclude the possibility of artifact results from contaminant phases. Temperature dependent magnetization M(T) curves under zero-field-cooled (ZFC) mode at a magnetic field of 1000 Oe are shown in Figure 6a and 6c. The ZFC curve at 1000 Oe magnetic field shows a distinct decrease in the temperature range from 2 to 70 K, which indicates that $V_2(Fe_xSn_{1-x})C$ exhibits a typical ferromagnetic-paramagnetic transition behavior. In the temperature range of 70-400 K, the ZFC curve has a more gradual decrease. Except at 2 K, all the magnetic hysteresis loops follow "S-shaped" character (Figure S6b) with small coercive force and residual magnetization (Table S8), suggesting that the $V_2(Fe_xSn_{1-x})C$ compound is a typical soft magnetic material (above 2 K) with saturation magnetization ($M_s$) gradually decreasing with increasing temperature. At 2 K, the coercive force ($H_c$), residual magnetization ($M_r$) and saturation magnetization ($M_s$) of $V_2(Fe_xSn_{1-x})C$ are 150.88 Oe, 0.00038 emu/g and 0.0806 emu/g, respectively.

In the case of $V_2(Fe_xCo_ySn_{1-x-y})C$, the ZFC curve (1000 Oe) shows a gradual drop of magnetization in the range 20-400 K (Figure 6c). The ferromagnetic-to-paramagnetic transition temperature is around 200 K according to the temperature-dependent measurements (Figure 5d). This temperature is much higher than for $V_2(Fe_xSn_{1-x})C$. The magnetic hysteresis loops were collected at 2, 50, 100, 200, 300 and 400 K, respectively (Figure 6d, Table S9). At 2 K, the coercive force ($H_c$), residual magnetization ($M_r$) and saturation magnetization ($M_s$) of $V_2(Fe_xCo_ySn_{1-x-y})C$ are 481.85 Oe, 0.0262 emu/g and 0.1378 emu/g, respectively.

Compared with Fe on A-sites of $V_2(Fe_xSn_{1-x})C$, the magnetization with two magnetic elements on A-sites of $V_2(Fe_xCo_ySn_{1-x-y})C$ is stronger. Therefore, it can be concluded that the magnetic properties can be tuned by adjusting the quantity and type of magnetic elements on the A sites. For an instance, the ZFC curve of $V_2(Fe_xCo_yNi_zSn_{1-x-y-z})C$ shows a gradual drop of magnetization at range of 20-400 K (Figure 6e). The ferromagnetic-to-paramagnetic transition temperature is around 200 K (Figure 6f). This temperature is much higher than that of $V_2(Fe_xSn_{1-x})C$ and close to that of $V_2(Fe_xCo_ySn_{1-x-y})C$. The residual magnetization ($M_r$) and saturation magnetization ($M_s$) at 2 K of $V_2(Fe_xCo_yNi_zSn_{1-x-y-z})C$ are 0.1499 emu/g and 0.7400 emu/g, respectively, much higher than for $V_2(Fe_xCo_ySn_{1-x-y})C$. In $V_2(Mn_xFe_yCo_zNi_nSn_{1-x-y-z-n})C$, which contains the antiferromagnetic element Mn, both the ZFC curve (1000 Oe) and the hysteresis loops shows changes (Figure 6g and 6h, Table S11). The ferromagnetic-to-paramagnetic transition temperature is shifted to almost 100 K. At 2 K, the coercive force ($H_c$), residual magnetization ($M_r$) and saturation magnetization ($M_s$) are 320.4 Oe, 0.0471 emu/g and 0.5677 emu/g, respectively. Although the ferromagnetic properties of $V_2(Mn_xFe_yCo_zNi_nSn_{1-x-y-z-n})C$ is less prominent than those of $V_2(Fe_xCo_yNi_zSn_{1-x-y-z})C$ in the range of 2-300 K, they are still much stronger than for the phases containing one or two magnetic elements. Thus, this approach provides a route for altering the magnetic properties of MAX phases by changing the chemical composition and component.

**Concluding remarks and outlook**

We have demonstrated that the series of $V_2(A'_xSn_{1-x})C$ phases (A'=Fe, Co, Ni and Mn, or their binary/ternary/quaternary combinations) can be synthesized by A-site element alloying, providing a generally applicable route to introduce one or more magnetic elements in A-sites and tune the resulting properties. A multielement phase $V_2(A'_xSn_{1-x})C$ (where again A' is a combination of Fe, Co, Ni and Mn, and x is close to 1/3) was realized. The fact that the magnetic properties are greatly enhanced for multielement A-layers lend credence to this concept and offers a rich chemical space for discovering new materials and properties using A-site multielement alloying strategy.

## Methods

### Materials

Elemental powders of vanadium (~300 mesh, 99.5 wt.% purity), tin (~300 mesh, 99.5 wt.% purity), iron (1 μm, 99.5 wt.% purity), cobalt (1 μm, 99.5% purity), nickel (1 μm, 99.5 wt.% purity), manganese (1 μm, 99.5 wt.% purity) and graphite (~300 mesh, 99.5 wt.% purity) were commercially obtained from Target Research Center of General Research Institute for Nonferrous Metals, Beijing, China. Sodium chloride (NaCl, 98 wt.%), potassium chloride (KCl, 98 wt.%), sulfuric acid ($H_2SO_4$, 98 wt.%)，hydrofluoric acid (HF, ≥40 wt.%), and absolute ethanol ($C_2H_6O$, 98 wt.%) were commercially obtained from Aladdin Chemical Reagent, China.

### Preparation of $V_2(A'_xSn_{1-x})C$ (A′ = Fe, Co, Ni, Mn, and combinations thereof)

The starting powders were mixed in a stoichiometric ratio of V:Sn:C:**A′**= 2:1.1:1:0.4 (the melting point of Sn was relatively low; we increased the content of tin because of the mass loss of tin at a high temperature), and the starting powders mixed with inorganic salt (NaCl+KCl). Moreover, for their binary/ternary/quaternary combination**,** the starting powders were mixed in a stoichiometric ratio, the starting powders mixed with inorganic salt (NaCl+KCl). The experimental details are shown in Table S12. After grinding for ten minutes, the starting powders mixed with inorganic salt (NaCl+KCl) were placed into an aluminum oxide boat. Then, the alumina boat was inserted into a tube furnace and heated to reaction temperature during 3 h with a heating rate of 10°C/min under an argon atmosphere. After the end of reaction, the products were washed, filtered, and dried at 40˚C in vacuum.

### Computational details

All the first-principles calculations were performed using the VASP code.[25,26] Based on the projector augmented wave (PAW) pseudopential[27] with a plane-wave cutoff energy of 500 eV, the generalized gradient approximation (GGA) as implemented by Perdew-Burke-Ernzerhof (PBE)[28] was employed for describing the exchange-correlation functional. All structures were relaxed until the force on each atom was less than $1.0\times10^{-3}$ eV, with the criterion for energy convergence set as $1.0\times10^{-6}$ eV. A Γ-centered k-point mesh of 12×12×3 was adopted for optimizing the unit cell of a 211 MAX phase. To investigate the solid solution, a 3×1×1 supercell of the 211 MAX phase was constructed, and an equivalent k-point density used for relaxation. Regarding atomic charges, a Bader charge analysis[29] based on 180×180×180 grid was performed. Phonon dispersion was investigated with Phonopy software[30] and the VASP code based on the density functional perturbation theory (DFPT).[31] A 3×3×1 supercell of the 211 MAX phase with a Γ-centered k-point mesh of 6×6×4 was employed for calculating the dynamical matrix. All the structures were visualized in the VESTA3 code.[32]

The CALPHAD approach was applied to calculate the phase diagrams of the V-Sn-C, V-Fe-C and V-Sn-Fe-C system. Due to lack of experimental data on the ternary V-Sn-C, V-Fe-C and V-Sn-Fe-C compounds, first-principles calculations were conducted to support the CALPHAD work.[33] The formation enthalpies of the $V_2SnC$ and $V_2FeC$ ternary compounds were computed to be -146.69 and -97.05 kJ/mol, respectively. The calculated formation enthalpies of $V_2(Sn,Fe)C$ show an approximately linear correlation versus the Fe content. Thus, the solid solution of $V_2SnC$ and $V_2FeC$ is considered to be an ideal solution in the present work. The Gibbs free energy function of $V_2(Sn,Fe)C$ was then determined with the Neumann-Kopp rule and added in the CALPHAD-type dataset of the V-Sn-Fe-C system, which included the thermodynamic parameters of the binary V-Sn, V-Fe, V-C, Sn-C, Sn-Fe and Fe-C systems.[34-38] With the established thermodynamic dataset, the isothermal sections of the V-Sn-C, V-Fe-C and V-Fe-Sn-C systems at 1100°C were computed. All the calculations were performed with the Thermo-Calc software.

**Characterization**

The phase composition of the samples was analyzed by X-ray diffraction (XRD, D8 Advance, Bruker AXS, Germany) with Cu Kα radiation. X-ray diffractograms were collected at a step size of 0.02° 2θ with a collection time of 1 s per step. The microstructure and chemical composition were observed by scanning electron microscopy (SEM, QUANTA 250 FEG, FEI, USA) equipped with an energy-dispersive spectrometer (EDS), and the atomic percentage values from EDS results were fitted by XPP. Structural and chemical analysis was carried out by high-resolution STEM high angle annular dark field (HRSTEM-HAADF) imaging and STEM affiliated energy dispersive X-ray (EDS) spectroscopy within Linköping's double Cs corrected FEI Titan3 60-300 microscope operated at 300 kV, and STEM-EDS was recorded with the embedded high sensitivity Super-X EDS detector.

The magnetic properties were measured on a Quantum Design superconducting quantum interference device magnetometer (SQUID). The powders were washed by sulfuric acid ($H_2SO_4$, 98 wt.%) and hydrofluoric acid (HF, ≥40 wt.%) to remove the Sn and Sn-containing intermetallic compounds. After washing, the powders were cold-pressed into round disks (diameter 10 mm) under a pressure of 20 MPa. A rectangular block of about 2 x 3 mm was cut the round disk, pasted on a quartz rod with tape and put into the SQUID instrument. Measurements were made in fields of 1000 Oe in the temperature range 2-400 K after cooling in zero applied field (ZFC) and in the measuring field (field-cooled, FC). The magnetic hysteresis loops measured in the fields of -10 kOe- 10 kOe at the temperature 2 K, 100 K, 200 K, 300 K, and 400 K, respectively.

**Data availability.** Essential all data generated or analyzed during this study are included in this published article (and its Supplementary Information files).

**ADDITIONAL INFORMATION**

Supplementary information is available in the online version of the paper. Correspondence and requests for materials should be addressed to Q. H., huangqing@nimte.ac.cn

**AUTHOR CONTRIBUTIONS**

Q. Huang initiated and supervised the work. Y. B. Li synthesized, analyzed, and characterized magnetic properties of the materials. J. Lu and P. O. Å. Persson carried out the STEM and analysis with input from P. Eklund and L. Hultman. S. Y. Du, X. H. Zha, K. Luo and K. Chang performed the theoretical part. M Li, Y. M Zhang, K. Chen, Z. R. Huang and Z. F. Chai contributed to the scientific discussion. Y.B.L. and Q.H. wrote the draft of the manuscript with contributions from P. E. K. K. Chang and X. H. Zha wrote the theory sections.

All coauthors contributed to the discussion and commented on successive versions of the manuscript.

†These authors contributed equally to this work.

**Conflict of interest:** The authors declare no competing financial interests.

**Acknowledgements**

  This study was supported financially by the National Natural Science Foundation of China (Grant No. 21671195, 21805295 and 21875271), Chinese Academy of Sciences (Grant No. 2019VEB0008 and 174433KYSB20190019). We acknowledge support from the Swedish Government Strategic Research Area in Materials Science on Functional Materials at Linköping University (Faculty Grant SFO‐Mat‐LiU No. 2009 00971). The Knut and Alice Wallenberg Foundation is acknowledged for support of the electron microscopy laboratory in Linköping (Grant KAW 2015.0043), an Academy Fellow Grant (P. E.) and a Scholar Grant (L. H.). P.O.Å.P. also acknowledges the Swedish Foundation for Strategic Research (SSF) through project funding (EM16‐0004) and the Research Infrastructure Fellow RIF 14‐0074.

# References

1. Baibich, M. N. et al. Giant magnetoresistance of (001)Fe/(001)Cr magnetic superlattices. *Phys. Rev. Lett.* **61,** 2472-2475 (1988).
2. Barsoum, M. W. The $M_{N+1}AX_N$ phases: A new class of solids: Thermodynamically stable nanolaminates. *Prog. Solid State Chem.* **28**, 201-281 (2000).
3. Sokol, M., Natu, V., Kota, S. & Barsoum, M. W. On the chemical diversity of the MAX Phases. *Trends in Chemistry* **1**, 210-223 (2019).
4. Ingason, A. S. et al. Magnetic self-organized atomic laminate from first principles and thin film synthesis. *Phys. Rev. Lett.* **110,** 195502 (2013).
5. Thore, A., Dahlqvist, M., Alling, B. & Rosen, J. Magnetic exchange interactions and critical temperature of the nanolaminate $Mn_2GaC$ from first-principles supercell methods. *Phys. Rev. B* **93,** 054432 (2016).
6. Tao, Q. et al. Thin film synthesis and characterization of a chemically ordered magnetic nanolaminate $(V,Mn)_3GaC_2$. *APL Mater.* 4, 086109 (2016).
7. Salikhov, R. et al. Magnetic properties of nanolaminated $(Mo_{0.5}Mn_{0.5})_2GaC$ MAX phase. *J. Appl. Phys.* **121,** 163904 (2017).
8. Petruhins, A. et al. Synthesis and characterization of magnetic $(Cr_{0.5}Mn_{0.5})_2GaC$ thin films. *J. Mater. Sci.* **50,** 4495-4502 (2015).
9. Lai, C.-C. et al. Magnetic properties and structural characterization of layered $(Cr_{0.5}Mn_{0.5})_2AuC$ synthesized by thermally induced substitutional reaction in $(Cr_{0.5}Mn_{0.5})_2GaC$. *APL Mater.* **6,** 026104 (2018).
10. Lin, S. et al. Alloying effects on structural, magnetic, and electrical/thermal transport properties in MAX-phase $Cr_{2-x}M_xGeC$ ( M = Ti, V, Mn, Fe, and Mo). *J. Alloys Compd.* **680,** 452-461 (2016).
11. Hamm, C. M., Bocarsly, J. D., Seward, G., Kramm, U. I. & Birkel, C. S. Non-conventional synthesis and magnetic properties of MAX phases $(Cr/Mn)_2AlC$ and $(Cr/Fe)_2AlC$. *J. Mater. Chem. C* **5,** 5700-5708 (2017).
12. Lai, C.-C. et al. Thermally induced substitutional reaction of Fe into $Mo_2GaC$ thin films. *Mater. Res. Lett.* **5,** 533-539 (2017).
13. Talapatra, A. et al. High-throughput combinatorial study of the effect of M site alloying on the solid solution behavior of $M_2AlC$ MAX phases. *Phys. Rev. B* **94,** 104106 (2016).
14. Ouabadi, N. et al. Formation mechanisms of $Ti_3SnC_2$ nanolaminate carbide using Fe as additive. *J. Am. Ceram. Soc.* **96,** 3239-3242 (2013).
15. Lapauw, T., Tunca, B., Cabioc'h, T., Vleugels, J. & Lambrinou, K. Reactive spark plasma sintering of $Ti_3SnC_2$, $Zr_3SnC_2$ and $Hf_3SnC_2$ using Fe, Co or Ni additives. *J. Eur. Ceram. Soc.* **37,** 4539-4545 (2017).
16. Tsai, M.-H. & Yeh, J.-W. High-entropy alloys: a critical review. *Mate. Res. Lett.* **2,** 107-123 (2014).
17. Miracle, D. B. & Senkov, O. N. A critical review of high entropy alloys and related concepts. *Acta Mater.* **122,** 448-511 (2017).
18. George, E. P., Raabe, D. & Ritchie, R. O. High-entropy alloys. *Nat. Rev. Mater.* **1,** (2019).
19. Tian, W.-B., Wang, P.-L., Kan, Y.-M. & Zhang, G.-J. $Cr_2AlC$ powders prepared by molten salt method. *J. Alloys Compd.* **461,** L5-L10 (2008).
20. Wang, B., Zhou, A., Hu, Q. & Wang, L. Synthesis and oxidation resistance of $V_2AlC$ powders by

|    | |
|----|---|
| | molten salt method. *Int. J. Appl. Ceram. Technol.* **14,** 873-879 (2017). |
| 21 | Eklund, P., Beckers, M., Jansson, U., Högberg, H. & Hultman, L. The $M_{n+1}AX_n$ phases: Materials science and thin-film processing. *Thin Solid Films* **518,** 1851-1878 (2010). |
| 22 | Fashandi, H. et al. Synthesis of $Ti_3AuC_2$, $Ti_3Au_2C_2$ and $Ti_3IrC_2$ by noble metal substitution reaction in $Ti_3SiC_2$ for high-temperature-stable Ohmic contacts to SiC. *Nat. Mater.* **16**, 814-818 (2017). |
| 23 | Li, M. et al. An Element Replacement Approach by Reaction with Lewis acidic Molten Salts to Synthesize Nanolaminated MAX Phases and MXenes. *J. Am. Chem. Soc.* **141,** 4730-4737 (2019). |
| 24 | Tsang, C., Heiman, N. & Lee, K. Exchange induced unidirectional anisotropy at FeMn-$Ni_{80}Fe_{20}$ interfaces. *J. Appl. Phys.* **52,** 2471-2473 (1981). |
| 25 | Kresse, G. & Furthmüller, J. Efficiency of ab-initio total energy calculations for metals and semiconductors using a plane-wave basis set. *Comput. Mater. Sci.* **6,** 15-50 (1996). |
| 26 | Kresse, G. & Furthmüller, J. Efficient iterative schemes for ab initio total-energy calculations using a plane-wave basis set. *Phys. Rev. B* **54,** 11169 (1996). |
| 27 | Blochl, P. E. Projector augmented-wave method. *J. Phys.: Condens. Matter* **50,** 17953-17979 (1994). |
| 28 | Perdew, J. P., Burke, K. & Ernzerhof, M. Generalized gradient approximation made simple. *Phys. Rev. Lett.* **77,** 3865 (1996). |
| 29 | Tang, W., Sanville, E. & Henkelman, G. A grid-based Bader analysis algorithm without lattice bias. *J. Phys.: Condens. Matter* **21,** 084204 (2009). |
| 30 | Togo, A., Oba, F. & Tanaka, I. First-principles calculations of the ferroelastic transition between rutile-type and $CaCl_2$-type $SiO_2$ at high pressures. *Phys. Rev. B* **78,** 134106 (2008). |
| 31 | Gonze, X. & Lee, C. Dynamical matrices, Born effective charges, dielectric permittivity tensors, and interatomic force constants from density-functional perturbation theory. *Phys. Rev. B* **55,** 10355 (1997). |
| 32 | Momma, K. & Izumi, F. VESTA 3for three-dimensional visualization of crystal, volumetric and morphology data. *J. Appl. Crystallogr.* **44,**1272-1276 (2011). |
| 33 | Chang, K., Hallstedt, B. & Music, D. Thermodynamic and Electrochemical Properties of the Li–Co–O and Li–Ni–O Systems. *Chem. Mater.* **24,** 97-105 (2011). |
| 34 | Kumar, K. H., Wollants, P. & Delaey, L. Thermodynamic evaluation of Fe-Sn phase diagram. *Calphad* **20,** 139-149 (1996). |
| 35 | Guo, C., Li, C., Zheng, X. & Du, Z. Thermodynamic modeling of the Fe–Ti–V system. *Calphad* **38,** 155-160 (2012). |
| 36 | Studnitzky, T., Onderka, B. & Schmid-Fetzer, R. Phase formation and reaction kinetics in the vanadium-tin system. *Z. Krist.-Cryst. Mater.* **93,** 48-57 (2002). |
| 37 | Chen, L. et al. Thermodynamic description of the Fe–Cu–C system. *Calphad* **64,** 225-235 (2019). |
| 38 | Huang, W. Thermodynamic properties of the Fe-Mn-VC system. *Metall. Trans. A* **22,** 1911-1920 (1991). |